\PassOptionsToPackage{fleqn}{amsmath}
\documentclass{easychair}

\usepackage[utf8]{inputenc}
\usepackage{microtype}%if unwanted, comment out or use option "draft"
\usepackage{amssymb}
\usepackage{tikz}
\usepackage{xspace}
\usepackage{subcaption}

\makeatletter

\newcommand{\r@rrow}[3]{%
  \newcommand{#1}[2][]{%
    \def\next{#2\@ifempty{##1}{}{_{##1}}\@ifempty{##2}{}{^{##2}}}%
    \mathchoice{#3[##1]{##2}}{\next}{\next}{\next}%
  }%
}
\newcommand{\l@rrow}[3]{%
  \newcommand{#1}[2][]{%
    \def\next####1{%
      \setbox0=\hbox{$####1\vphantom{#2}\@ifempty{##1}{}{_{\vphantom{##1}}}%
      \@ifempty{##2}{}{^{##2}}$}%
      \setbox1=\hbox{$####1\vphantom{#2}\@ifempty{##1}{}{_{##1}}%
      \@ifempty{##2}{}{^{\vphantom{##2}}}$}%
      \setbox2=\vbox{\hbox to\wd0{}\hbox to\wd1{}}%
      \mathrel{\hskip\wd2\hskip-\wd0\box0\hskip-\wd1\box1{#2}}%
    }%
    \mathchoice{#3[##1]{##2}}{\next\textstyle}%
    {\next\scriptstyle}{\next\scriptscriptstyle}%
  }%
}

\l@rrow{\xl}{\leftarrow}{\xleftarrow}
\r@rrow{\xr}{\rightarrow}{\xrightarrow}
\l@rrow{\xphl}{\phleftarrow}{\xphleftarrow}
\r@rrow{\xphr}{\phrightarrow}{\xphrightarrow}

\makeatother 

\title{%
Formalized Confluence of Quasi-Decreasing,\\ Strongly Deterministic Conditional TRSs%
\footnote{%
  The research described in this paper is supported by FWF (Austrian
  Science Fund) project P27502.}}

\author{
  Thomas Sternagel
\and
  Christian Sternagel
}

\institute{
  University of Innsbruck, Austria\\
  \email{\{thomas,christian\}.sternagel@uibk.ac.at}
 }

\authorrunning{T.~Sternagel and C.~Sternagel}

\titlerunning{Conditional Critical Pair Criterion Formalized}

\theoremstyle{plain}

\let\oldeqref\eqref
\renewcommand\eqref[1]{\oldeqref{eq:#1}}

\def\systemname#1{\mbox{\textsf{#1}}\xspace}

\newcommand{\concon}{\systemname{ConCon}}
\newcommand\ceta{\textsf{C\kern-0.2exe\kern-0.5exT\kern-0.5exA}\xspace}

\newcommand{\TTTT}{%
 \systemname{T\kern-0.2em\raisebox{-0.3em}T\kern-0.2emT\kern-0.2em%
 \raisebox{-0.3em}2}%
}

\newcommand\mc[1]{\ensuremath{\mathcal{#1}}\xspace}
\newcommand\msf[1]{\ensuremath{\mathsf{#1}}\xspace}
\newcommand\FF{\mc{F}}
\newcommand\VV{\mc{V}}

\newcommand\RR{\mc{R}}
\newcommand\TT{\mc{T}}
\newcommand\EE{\mc{E}}
\newcommand{\Evars}{\ensuremath{\EE\VV}}
\newcommand\subt{\msf{st}}

\newcommand\sig{\FF}
\newcommand\vars{\VV}

\newcommand\termsover[2]{\ensuremath{\TT(#1,#2)}}
\newcommand\terms{\termsover{\sig}{\vars}}

\newcommand\IF{\Leftarrow}
\newcommand\crule[1]{\ell_{#1} \to r_{#1} \IF c_{#1}}
\newcommand{\subterm}{\ensuremath{\mathrel{\vartriangleright}}}

\newcommand\unifsymbol{\sim}
\newcommand\unif{\mathrel{\unifsymbol}}

\newcommand\cond{\approx}
\newcommand\isafor{\textsf{Isa\kern-0.2exF\kern-0.2exo\kern-0.2exR}\xspace}
\newcommand\isaforchangeset{dbc03280d673}
\newcommand\isaforpath{http://cl2-informatik.uibk.ac.at/rewriting/%
  mercurial.cgi/IsaFoR/file/\isaforchangeset/thys}
\newcommand\thyref[1]{\href{\isaforpath/#1.thy}{\nolinkurl{#1}}}
\newcommand\factref[3]{\href{\isaforpath/#2.thy\#l#3}{\nolinkurl{#1}}}
\newcommand\wrt{with respect to\xspace}
\newcommand\papp{}
\newcommand\sigmaren{\pi^-\sigma_1}

\newcounter{counter}

\theoremstyle{definition}

\theoremstyle{plain}
\newtheorem{theorem}[counter]{Theorem}

\theoremstyle{remark}

\begin{document}

\maketitle

\pagestyle{empty}

%------------------------------------------------------------------------------
\begin{abstract}
We present an Isabelle/HOL formalization
of a characterization of confluence for quasi-reductive strongly deterministic
conditional term rewrite systems,
due to Avenhaus and Loría-Sáenz.
\end{abstract}

\section{Introduction}
\label{sect:introduction}

Already in 1994 Avenhaus and Loría-Sáenz~\cite{AL94} proved a critical
pair criterion for deterministic conditional term rewrite systems with extra
variables in right-hand sides, provided their rewrite relation is decidable
and terminating.
We use this criterion in our conditional confluence checker \concon~\cite{SM14}.
In the following we provide a description of our formalization of the
conditional critical pair criterion where we strengthened the original result
from quasi-reductivity to quasi-decreasingness.
This is a first step towards certifying the confluence criterion that a
quasi-decreasing and strongly deterministic CTRS is confluent if all of its
critical pairs are joinable.
The formalization described in this paper is part of a greater effort to
formalize all methods employed by \concon to be able to certify its output.

\subparagraph{Contribution.}
We have formalized Theorem~4.1 from Avenhaus and Loría-Sáenz~\cite{AL94}
in Isabelle/HOL~\cite{Isabelle} as well as strengthened the original theorem
from quasi-reductivity to quasi-decreasingness. It is now part of the formal library \isafor~\cite{TS09} (the Isabelle
Formalization of Rewriting) and freely available online at:
\begin{quote}\small
\url{\isaforpath/Conditional_Rewriting/ALS94.thy}
\end{quote}

\section{Preliminaries}

We assume familiarity with the basic notions of (conditional) term
rewriting~\cite{BN98,O02}, but shortly recapitulate terminology and notation
that we use in the remainder.
%relations
Given an arbitrary binary relation $\xr[\alpha]{}$,
we write $\xl[\alpha]{}$, $\xr[\alpha]{+}$, $\xr[\alpha]{*}$ for the
\emph{inverse},
the \emph{transitive closure}, and the \emph{reflexive transitive closure}
of $\xr[\alpha]{}$,
respectively.
% variables
We use $\vars(\cdot)$ to denote the set of variables occurring in a given
syntactic object, like a term, a pair of terms, a list of terms, etc.
% terms
The set of terms $\TT(\FF,\VV)$ over a given signature of function symbols $\FF$
and set of variables $\VV$ is defined inductively:
$x \in \TT(\FF, \VV)$ for all variables $x \in \VV$,
and for every $n$-ary function symbol $f \in \FF$ and terms $t_1,\ldots,t_n \in
\TT(\FF,\VV)$ also $f(t_1,\ldots,t_n) \in \TT(\FF,\VV)$.
% unification
We say that terms $s$ and $t$ \emph{unify}, written $s \unif t$, if $s\sigma =
t\sigma$ for some substitution~$\sigma$.
% substitutions
A substitution $\sigma$ is \emph{normalized \wrt $\RR$} if $\sigma(x)$ is a
normal form \wrt $\xr[\RR]{}$ for all $x \in \VV$.
We call a bijective variable substitution $\pi : \VV \to \VV$ a \emph{variable
renaming} or \emph{(variable) permutation}, and denote its inverse by $\pi^-$.
% strongly irreducible term
A term $t$ is \emph{strongly irreducible \wrt $\RR$} if $t\sigma$ is a normal
form \wrt $\xr[\RR]{}$ for all normalized substitutions $\sigma$.
% SDTRS
A \emph{strongly deterministic oriented 3-CTRS (SDTRS)} $\RR$ is a set of conditional
rewrite rules of the shape $\crule{}$ where $\ell$ and $r$ are terms and $c$ is
a possibly empty sequence of pairs of terms $s_1 \approx t_1, \ldots, s_n
\approx t_n$.
For all rules in $\RR$ we have that
$\ell \not\in \VV$,
$\vars(r) \subseteq \vars(\ell,c)$, 
$\vars(s_i) \subseteq \vars(\ell,t_1,\ldots,t_{i-1})$
for all $1 \leqslant i \leqslant n$, and
$t_i$ is strongly irreducible \wrt $\RR$ for all $1 \leqslant i \leqslant n$.
% labeled rules
We sometimes label rules like $\rho: \crule{}$.
% extra variables
For a rule $\rho: \crule{}$ of an SDTRS $\RR$ the set of \emph{extra variables}
is defined as $\Evars(\rho) = \vars(c) - \vars(\ell)$.
% rewrite relation
The rewrite relation $\xr[\RR]{}$ is the smallest relation $\xr[]{}$ satisfying
$t[\ell\sigma]_p \xr[]{} t[r\sigma]_p$ whenever $\crule{}$ is a rule in $\RR$
and $s\sigma \xr[\RR]{*} t\sigma$ for all $s\cond t \in c$.
% conditional overlaps
Two variable-disjoint variants of rules $\crule{1}$ and $\crule{2}$ in
$\RR$ such that $\ell_1|_p \notin \VV$ and $\ell_1|_p\mu = \ell_2\mu$ with most
general unifier (mgu)~$\mu$, constitute a \emph{conditional overlap}.
% conditional critical pairs
A conditional overlap that does not result from overlapping two variants of the
same rule at the root, gives rise to a \emph{conditional critical pair} (CCP)
$r_1\mu \cond r_1[r_2]_p\mu \IF c_1\mu,c_2\mu$.
% joinability of CCPs
A CCP $u \cond v \IF c$ is \emph{joinable} if $u\sigma \downarrow_\RR v\sigma$ for
all substitutions $\sigma$ such that $s\sigma \xr[\RR]{*} t\sigma$ for all $s
\cond t \in c$.
% ordering
We denote the proper subterm relation by $\subterm$ and define
${\succ_\subt}={(\succ \cup \subterm)^+}$ for some reduction order $\succ$.
% quasi-reductivity
Let $\succ$ be a reduction order on $\TT(\FF,\VV)$ then an SDTRS $\RR$ is
\emph{quasi-reductive \wrt $\succ$} if for every substitution $\sigma$ and every
rule $\ell \to r \IF s_1 \cond t_1, \ldots, s_n \cond t_n$ in $\RR$ we have
$s_j\sigma \succeq t_j\sigma$ for $1 \leqslant j \leqslant i$ implies $\ell\sigma
\succ_\subt s_{i+1}\sigma$, and
$s_j\sigma \succeq t_j\sigma$ for $1 \leqslant j \leqslant n$ implies $\ell\sigma
\succ r\sigma$.%
\footnote{%
  This is the definition from \cite{AL94} which differs from the one in
  \cite[Definition~7.2.36]{O02} in two respects. First $\succ$ is a reduction
  order (hence also closed under substitutions; this is needed in the proof of
  \cite[Theorem 4.2]{AL94}) whereas in Ohlebusch $\succ$ is
  a well-founded partial order that is closed under contexts.
  Moreover Ohlebusch allows a signature extension for the substitutions $\sigma$
  which is not part of this definition.
}
% quasi-decreasingness
On the other hand, an SDTRS $\RR$ over signature $\FF$ is \emph{quasi-decreasing}
if there is a well-founded order $\succ$ on $\terms$ such that
${\succ} = {\succ_\subt}$,
${\xr[\RR]{}} \subseteq {\succ}$,
and for all rules 
$\ell\to r\Leftarrow s_1\approx t_1,\ldots,s_n\approx t_n$ in $\RR$,
all substitutions $\sigma \colon \VV \to \terms$, and
$0\leqslant i < n$, if
$s_j\sigma \xr[\RR]{*} t_j \sigma$
for all $1 \leqslant j \leqslant i$
then $\ell \sigma \succ s_\text{i+1} \sigma$
.
Quasi-reductivity implies quasi-decreasingness (cf.~\cite[proof of
Lemma~7.2.40]{O02}).

\section{Confluence of Quasi-Decreasing SDTRSs}

The main result from Avenhaus and Loría-Sáenz is the following theorem:

\begin{theorem}[{Avenhaus and Loría-Sáenz~\cite[{Theorem~4.1}]{AL94}}]
  Let $\RR$ be an SDTRS that is quasi-reductive \wrt $\succ$. $\RR$ is
  confluent if and only if all conditional critical pairs are joinable.
\end{theorem}

That all critical pairs of any CTRS $\RR$ (no need for strong determinism or
quasi-reductivity) are joinable if $\RR$ is confluent is straight-forward so we
will concentrate on the other direction. Our formalization is quite close to
the original proof. The good news is: we could not find any
errors (besides typos) in the original proof but as is often the case with
formalizations there are places where the paper proof is too vague or does not
spell out the technical details in favor of readability. A luxury we cannot
afford. For example we heavily rely on an earlier formalization
of permutations \cite{HMS14} in order to formalize variants of
rules up to renaming.
Even the change from quasi-reductivity to quasi-decreasingness did not pose a
problem.

In the following we will give a description of
the main theorem of our formalization and its proof.

\begin{theorem}
Let $\RR$ be an SDTRS that is quasi-decreasing \wrt $\succ$ and where all
conditional critical pairs are joinable, then $\RR$ is confluent.
\end{theorem}
\begin{figure}[h!]
  \centering
  \begin{subfigure}[t]{0.5\textwidth}
  \centering
  \begin{tikzpicture}[node distance=1.3cm]
    \node (s) at (0,0) {$s$};
    \node (t1) [below left of=s] {$t'$};
    \node (t2) [below right of=s] {$u'$};
    \node (m) [below right of=t1] {$\cdot$};
    \node (t) [below left of=t1] {$t$};
    \node (u) [below right of=t2] {$u$};
    \node (s') [below right of=t] {$\cdot$};
    \node (s'') [below right of=s'] {$\cdot$};
    \node (IH) [below of=t1,yshift=4mm] {IH};
    \node (IH2) [below right of=m,yshift=4mm,xshift=-5mm] {IH};

    \draw[->] (s) -- node [sloped,above] {$\mathrel{}_\subt\prec$} (t1);
    \draw[->] (s) -- node [sloped,above] {$\succ_\subt$} (t2);
    \draw[->] (t1) -- node[sloped,above] {$\scriptstyle *$} (t);
    \draw[->] (t2) -- node[sloped,above] {$\scriptstyle *$} (u);
    \draw[->,dotted] (t) -- node[sloped,above] {$\scriptstyle *$} (s');
    \draw[->,dotted] (s') -- node[sloped,above] {$\scriptstyle *$} (s'');
    \draw[->,dotted] (u) -- node[sloped,above] {$\scriptstyle *$} (s'');
    \draw[->] (t1) -- node[sloped,above] {$\scriptstyle *$} (m);
    \draw[->] (t2) -- node[sloped,above] {$\scriptstyle *$} (m);
    \draw[->,dotted] (m) -- node[sloped,above] {$\scriptstyle *$} (s');
  \end{tikzpicture}
  \caption{}
  \label{fig:proof}
  \end{subfigure}%
  \begin{subfigure}[t]{0.5\textwidth}
  \centering
  \begin{tikzpicture}[node distance=1.3cm]
    \node (0) at (0,0) {$t_{i+1}\sigma'_1$};
    \node (1) [above right of=0] {$s_{i+1}\sigmaren$};
    \node (2) [below right of=1] {$s_{i+1}\sigma'_1$};
    \node (3) [above right of=2] {$s_{i+1}\sigma_2$};
    \node (4) [below right of=3] {$t_{i+1}\sigma'_2$};
    \node (5) [below right of=0] {$\cdot$};
    \node (6) [below right of=5] {$\cdot$};
    \node (IH) [below of=1,yshift=4mm] {IH};
    \node (IH2) [below right of=2,yshift=4mm,xshift=-5mm] {IH};

    \draw[->] (1) -- node[sloped,above] {$\scriptstyle *$} (0);
    \draw[->] (1) -- node[sloped,above] {$\scriptstyle *$} (2);
    \draw[->] (3) -- node[sloped,above] {$\scriptstyle *$} (2);
    \draw[->] (3) -- node[sloped,above] {$\scriptstyle *$} (4);
    \draw[->,dotted] (0) -- node[sloped,above] {$\scriptstyle *$} (5);
    \draw[->,dotted] (2) -- node[sloped,above] {$\scriptstyle *$} (5);
    \draw[->,dotted] (5) -- node[sloped,above] {$\scriptstyle *$} (6);
    \draw[->,dotted] (4) -- node[sloped,above] {$\scriptstyle *$} (6);
  \end{tikzpicture}
  \caption{}
  \label{fig:proof2}
  \end{subfigure}
  \caption{}
\end{figure}
\begin{proof}
  Assume that all critical pairs are joinable.
  We will look at an arbitrary peak $t \xl[\RR]{*} s \xr[\RR]{*} u$ and prove
  that $t \downarrow_\RR u$ by well-founded induction on the relation
  $\succ_\subt$.
  If $s = t$ or $s = u$ then $t$ and $u$ are trivially joinable and we are
  done. So we may assume that the peak contains at least one step in each
  direction: $t \xl[\RR]{*} t' \xl[\RR]{} s \xr[\RR]{} u' \xr[\RR]{*} u$.

  We will proceed to prove that $t' \downarrow_\RR u'$ then $t \downarrow_\RR
  u$ follows by two applications of the induction hypothesis as shown in
  Figure~\ref{fig:proof}.
  Assume that $s = C[\ell_1\sigma_1]_p \xr[\RR]{} C[r_1\sigma_1]_p = t'$
  and $s = D[\ell_2\sigma_2]_q \xr[\RR]{} D[r_2\sigma_2]_q = u'$ for rules
  $\rho_1:\crule{1}$ and $\rho_2:\crule{2}$ in $\RR$, contexts $C$ and $D$,
  positions $p$ and $q$, and substitutions $\sigma_1$ and $\sigma_2$
  such that $u\sigma_1 \xr[\RR]{*} v\sigma_1$ for all $u \cond v \in c_1$
  and $u\sigma_2 \xr[\RR]{*} v\sigma_2$ for all $u \cond v \in c_2$.
  There are three possibilities: $p \parallel q$, $p \leqslant q$, or $q
  \leqslant p$.
  In the first case $t' \downarrow_\RR u'$ holds because the two
  redexes do not interfere. The other two cases are symmetric and we only
  consider $p \leqslant q$ here.
  If ${s}\subterm{s|_p}={\ell_1\sigma_1}$ then ${s}\succ_\subt{\ell_1\sigma_1}$ (by
  definition of $\succ_\subt$) and there is a position $r$ such that $q = p\papp r$
  and so we have the peak $r_1\sigma_1 \xl[\RR]{*} \ell_1\sigma_1 \xr[\RR]{*}
  \ell_1\sigma_1[r_2\sigma_2]_r$ which is joinable by induction hypothesis.
  But then the peak $t' = s[r_1\sigma_1]_p \xl[\RR]{*} s[\ell_1\sigma_1]_p
  \xr[\RR]{*} s[\ell_1\sigma_1[r_2\sigma_2]_r]_q = u'$ is also joinable (by
  closure under contexts) and we are done.
  So we may assume that $p = \epsilon$ and thus $s = \ell_1\sigma_1$.
  Now, either $q$ is a function position in $\ell_1$ or there is
  a variable position $q'$ in $\ell_1$ such that $q' \leqslant q$.
  In the first case we either have a CCP which is joinable by assumption
  or we have a root-overlap of variants of the same rule.
  Then $\rho_1\pi = \rho_2$ for some permutation $\pi$. Moreover,
  $s = \ell_1\sigma_1 = \ell_2\sigma_2$ and we have
  \begin{equation}
  x\sigmaren =
  x\sigma_2 \text{ for all variables $x$ in
  $\vars(\ell_2)$.}\label{eq:vars}
  \end{equation}
  We will prove $x\sigmaren \downarrow_\RR x\sigma_2$ for all $x$ in
  $\vars(\rho_2)$.
  Since $t' = r_1\sigma_1 = r_2\sigmaren$ and $u' = r_2\sigma_2$ this shows $t'
  \downarrow_\RR u'$.
  Because $\RR$ is terminating (by quasi-decreasingness) we may define two
  normalized substitutions $\sigma'_i$ such that
  \begin{equation}
  x\sigmaren \xr[\RR]{*} x\sigma'_1
  \text{ and }
  x\sigma_2 \xr[\RR]{*} x\sigma'_2
  \text{ for all variables $x$.}
  \label{eq:vars'}
  \end{equation}
  We prove $x\sigma'_1 = x\sigma'_2$ for $x \in \Evars(\rho_2)$ by an inner induction
  on the length of $c_2 = s_1 \cond t_1, \ldots, s_n \cond t_n$.
  If $\rho_2$ has no conditions this holds vacuously
  because there are no extra variables.
  In the step case the inner induction hypothesis is that
  $x\sigma'_1 = x\sigma'_2$ for $x \in
  \vars(s_1,t_1,\ldots,s_i,t_i)-\vars(\ell_2)$
  and we have to show that
  $x\sigma'_1 = x\sigma'_2$ for $x \in
  \vars(s_1,t_1,\ldots,s_{i+1},t_{i+1})-\vars(\ell_2)$.
  If $x \in \vars(s_1,t_1,\ldots,s_i,t_i,s_{i+1})$ we are done by the inner induction
  hypothesis and strong determinism of $\RR$.
  So assume $x \in \vars(t_{i+1})$.
  From strong determinism of $\RR$, \eqref{vars}, \eqref{vars'}, and the induction hypothesis we have
  that $y\sigma'_1 = y\sigma'_2$ for all $y\in\vars(s_{i+1})$
  and thus $s_{i+1}\sigma'_1 = s_{i+1}\sigma'_2$.
  With this we can find a join between $t_{i+1}\sigma'_1$ and $t_{i+1}\sigma'_2$
  by applying the induction hypothesis twice as shown in Figure~\ref{fig:proof2}.
  Since $t_{i+1}$ is strongly irreducible and $\sigma'_1$ and $\sigma'_2$ are
  normalized, this yields $t_{i+1}\sigma'_1 = t_{i+1}\sigma'_2$
  and thus $x\sigma'_1 = x\sigma'_2$.
  
  We are left with the case that there is a variable position $q'$ in $\ell_1$
  such that $q = q'r'$ for some position $r'$. Let $x$ be the variable $\ell_1|_{q'}$. Then
  $x\sigma_1|_{r'} = \ell_2\sigma_2$, which implies
  $x\sigma_1 \xr[\RR]{*} x\sigma_1[r_2\sigma_2]_{r'}$.
  Now let $\tau$ be the substitution such that $\tau(x) =
  x\sigma_1[r_2\sigma_2]_{r'}$ and
  $\tau(y) = \sigma_1(y)$ for all $y \neq x$,
  and $\tau'$ some normalization, i.e.,
  $y\tau \xr[\RR]{*} y\tau'$ for all $y$.
  Moreover, note that
  \begin{equation}
  y\sigma_1 \xr[\RR]{*} y\tau \text{ for all $y$.}
  \label{eq:3}
  \end{equation}
  We have $u' = \ell_1\sigma_1[r_2\sigma_2]_q = \ell_1\sigma_1[x\tau]_{q'}
  \xr[\RR]{*} \ell_1\tau$, and thus
  $u' \xr[\RR]{*} \ell_1\tau'$.
  From \eqref{3} we have $r_1\sigma_1 \xr[\RR]{*} r_1\tau$ and thus
  $t' = r_1\sigma_1 \xr[\RR]{*} r_1\tau'$.
  Finally, we will show that $\ell_1\tau' \xr[\RR]{} r_1\tau'$, concluding the
  proof of $t' \downarrow_\RR u'$.
  To this end, let $s_i \cond t_i \in c_1$. By \eqref{3} and the definition of
  $\tau'$ we obtain  $s_i\sigma_1 \xr[\RR]{*}
  t_i\sigma_1 \xr[\RR]{*} t_i\tau'$ and $s_i\sigma_1 \xr[\RR]{*} s_i\tau'$.
  But then, by induction hypothesis, $s_i\tau' \downarrow_\RR t_i\tau'$, and
  furthermore, since $t_i$ is strongly irreducible, $s_i\tau' \xr[\RR]{*}
  t_i\tau'$.
\end{proof}

\section{Conclusion}

Our formalization amounts to approximately 1800 lines of Isabelle. At some
points we actually had to use variants of rules where the original proof assumes
two rules to be identical. Apart from that the formalization was rather
straight-forward. Also the modification from quasi-reductivity to
quasi-decreasingness did not pose a problem.

\subparagraph*{Future Work.}
Formalizing the conditional critical pair criterion was only the first step.
There are two challenges for automation: Checking if a term is strongly
irreducible, and checking if a conditional critical pair is joinable. Both of
these are undecidable in general.
Avenhaus and Loría-Sáenz employ \emph{absolute determinism}
\cite[Definition~4.2]{AL94} to tackle
strong irreducibility as well as \emph{contextual rewriting} to handle
joinability of conditional critical pairs. Then we have a computable
overapproximation.
We already started to extend our formalization to facilitate absolute
determinism as well as contextual rewriting.
It remains to provide check functions for \ceta~\cite{TS09} and also the proper
certifiable output for \concon.

\subparagraph*{Acknowledgments.}

We thank the Austrian Science Fund (FWF project P27502) for supporting
our work.
Moreover we would like to thank the anonymous reviewers for useful suggestions.

\label{sect:bib}
\bibliography{references}

%------------------------------------------------------------------------------
\end{document}